\documentclass[aip,graphicx,preprint]{revtex4}

\draft 

\usepackage{amsmath}

\newcommand{\rtta}[1]{\bar{\bar{#1}}}
\usepackage{graphicx}
\begin{document}


\title{Optical activity of oriented molecular systems in terms of the magnetoelectric tensor of gyrotropy} 



\author{Oriol Arteaga}
\affiliation{Dep. F\'{i}sica Aplicada i \`{O}ptica, IN2UB, Barcelona, SPAIN}


\date{\today}

\begin{abstract}
The optical activity of oriented molecular systems is investigated using bianisotropic material constitutives for Maxwell's equations. It is shown that the circular birefringence and circular dichroism for an oriented system can be conveniently expressed in terms of the two components of the symmetric magnetoelectric tensor of gyrotropy that are perpendicular to this direction of light propagation. This description establishes a direct link between measurable anisotropic optical activity and the tensors that describe the oscillating electric and magnetic dipole and electric quadrupole moments induced by the optical wave.
\end{abstract}


\maketitle 


\section{Introduction}

The different response of a molecule to right- and left- circularly polarized light can be measured experimentally and can be explained with a semiclassical theory of multipole oscillators induced by the electric and magnetic fields of the optical wave \cite{book2}.  Thanks to advances in computational chemistry calculations of  the optical activity of oriented molecules are today relatively easy to do using modern software packages that do quantum-mechanical calculations of molecular property tensors \cite{g09}. Measurements of optical activity started more than two centuries ago and modern chiroptical spectroscopy techniques that exploit the intrinsic chirality of circular polarized light are widely available in chemistry laboratories. However, the experimental study of the optical activity of oriented systems is still challenging, and active research is being carried to find reliable methods to measure the typically small chiroptical contributions embedded in the large optical anisotropy of oriented system. In last few years Mueller matrix spectroscopy has emerged as a promising and powerful technique for these type of measurements \cite{tesismia}. Therefore, it seems that we have reached a point where the experimental and theoretical approaches to study the optical activity of oriented molecules are mature enough to be compared. One additional factor that complicates the reconciliation between these two approaches is that theoretical calculations of optical activity are typically done assuming single molecules while, in most cases, measurements are performed in macroscopic media.

Traditionally, measurements of ensembles of molecules were always done in solution, in which small molecules tend to adopt random orientations. Therefore calculations were always made under the assumption of a large collections of molecules randomly oriented. However, in recent years there is a growing interest to study the richer anisotropic spectroscopic information provided by oriented molecules. There are many strategies to orient a molecule, specially if the molecules are large. The most evident strategy consist of crystallizing them and form a molecular crystal but, in some occasions, it is also possible to incorporate the molecules during the growth of a different crystal that acts as a host \cite{bartdye}. In these cases the optical properties not only depend on the individual molecules but also in the type of crystal lattice. Elongated molecular aggregates are typically orientable by flows \cite{flow} and some molecules can also be oriented by simple mechanical actions such as rubbing \cite{wong}. Other methods to orient molecules use lasers pulses or other forms of electric and magnetic fields \cite{gueri}. However, in this work, we will consider that there are no additional external fields affecting the molecules during the optical activity measurements.

 In this paper we discuss the theoretical basis that permits the correlation between computations of the averaged molecular polarizability tensors and measurements. We show that the mathematical tools used to describe measurements and calculations naturally converge if a bianisotropic formulation of the material constitutive equations is considered. The key element of the analysis is to refer any measured or calculated optical activity to elements of the magnetoelectric tensor of gyration that is included in the bianisotropic constitutive equations. The results we find simplify the comparison between calculations and experiments.

\section{Bianisotropic constitutive equations and multipole theory}

The following form of bianisotropic constitutive equations was first given by Tellegen \cite{tellegen}:
\begin{subequations}\label{uno}
\begin{equation}
\mathbf{D}=\rtta{\varepsilon} \mathbf{E}+\rtta{\rho} \mathbf{H},
\end{equation}
\begin{equation}
\mathbf{B}=\rtta{\mu} \mathbf{H}+\rtta{\rho'} \mathbf{E},
\end{equation}
\end{subequations}
where frequency-domain fields are considered and the choice of the time dependence is given by $\exp(i\omega t)$. $\rtta{\varepsilon}$ is the permittivity dyadic, $\rtta{\mu}$ the permeability dyadic and $\rtta{\rho}$ and $\rtta{\rho'}$ are the two magnetoelectric dyadics \cite{malaka} and transmit the relation between the electric and magnetic field quantities $\mathbf{E}$, $\mathbf{H}$ and the flux quantities $\mathbf{D}$, $\mathbf{B}$. These four constitutive dyadics contain full information of the electromagnetic response of a bianisotropic medium. Lately, this form about the constitutive equations has attracted a lot of attention for the optical characterization of metamaterials \cite{advances}, because these materials typically have large magnetoelectric tensors.  One common simplification for Eqs. \eqref{uno} is that the specific medium should be Lorentz-reciprocal \cite{lakabook1}. This implies:
\begin{equation}\label{const}
\rtta{\varepsilon}=\rtta{\varepsilon}^T, \quad \rtta{\mu}=\rtta{\mu}^T, \quad \rtta{\rho}=-\rtta{\rho}'^T,
\end{equation}
where the superscript $T$ indicates transposition. In this case the magnetoelectric dyadics, that in general are written as $\rtta{\rho}=\rtta{\chi}+i\rtta{\kappa}$ and $\rtta{\rho'}=\rtta{\chi}^T-i\rtta{\kappa}^T$, must satisfy $\rtta{\chi}=0$ and $\rtta{\kappa}\neq0$. Therefore, the constitutive equations for a bianisotropic reciprocal medium can be rewritten as:
\begin{subequations}\label{constitutivesimplied2}
\begin{equation}
\mathbf{D}=\rtta{\varepsilon} \mathbf{E}+i\rtta{\kappa} \mathbf{H},
\end{equation}
\begin{equation}
\mathbf{B}=\rtta{\mu} \mathbf{H}-i\rtta{\kappa}^T \mathbf{E}.
\end{equation}
\end{subequations}

It should be indicated that, in the presence of absorption, all the tensors in this constitutive equations, i.e. $\rtta{\varepsilon}$, $\rtta{\kappa}$ and $\rtta{\mu}$, become complex. For systems uni- and biaxial orthorhombic crystallographic symmetries the real and imaginary parts of the tensor have the same system of principal axes. However, the situation is more complicated if the tensors have no such system, as can happen for media with monoclinic and triclinic symmetries, as they can lead to apparent non-reciprocal optical response, despite being reciprocal media.

Multipole theory can be used to calculate the reciprocal magnetolectric coupling (now given by $\rtta{\kappa}$) between the electric and magnetic fields \cite{raab2}. Note that Eqs. \eqref{const} do not necessarily constraint $\rtta{\kappa}$ to be a symmetric tensor. However, natural optical activity is a Lorentz reciprocal effect and only it can be contributed by the symmetric part of the magnetoelectric tensor. In this work we only consider this part of the tensor as we are interested in natural optical activity and non-magnetic media. For a complete description of natural optical activity it has been shown that,  within a semiclassical theory,  in addition to the mean electric dipole induced by the electric field of light it is necessary to include: the electric dipole contribution by the time-derivative of the magnetic field of the light wave, the associated mean magnetic dipole induced by the time-derivative of the electric field, the electric dipole contribution induced by the electric field gradient of the electromagnetic wave and the electric quadrupole contribution induced by the electric field \cite{dunn}. When all these contributions are considered the oscillating induced moments (the electric dipole $\hat{\mu}_{\alpha}$, the magnetic dipole $\hat{m}_{\alpha}$ and the electric quadrupole $\hat{\Theta}_{\alpha\beta}$) are the real parts of the following complex expressions \cite{book2}:
\begin{subequations}\label{moments}
\begin{equation}
\hat{\mu}_\alpha=\hat{\alpha}_{\alpha\beta}\hat{E}_{\beta}+\hat{G}_{\alpha\beta}\hat{B}_{\beta}+\frac{1}{3}\hat{A}_{\alpha\beta\gamma}\nabla_{\beta}\hat{E}_{\gamma}+...,
\end{equation}
\begin{equation}
\hat{m}_\alpha=\hat{G}_{\beta\alpha}\hat{E}_{\beta}+...,
\end{equation}
\begin{equation}
\hat{\Theta}_{\alpha\beta}=\hat{A}_{\gamma\alpha\beta}\hat{E}_{\gamma}+...,
\end{equation}
\end{subequations}
where $\hat{E}_{\alpha}$ and $\hat{B}_{\alpha}$ the actual electric and magnetic fields of the optical wave and $\nabla_{\beta}\hat{E}_{\gamma}$ is the electric field gradient. $\hat{\alpha}_{\alpha\beta}$ is the electric dipole-electric dipole polarizability tensor, $\hat{G}_{\alpha\beta}$ is the electric dipole-magnetic dipole polarizability tensor and $\hat{A}_{\alpha\beta\gamma}$ is the electric dipole-electric quadrupole polarizability tensor. Higher order polarizabilities (e. g. electric octopole, magnetic quadrupole,...) have not been specified because they do not contribute to optical activity \cite{raabbook}. The hats  $\hat{~}$  stress that these are complex quantities. Computational software has been developed to compute all these tensors quantum mechanically by applying time-dependent perturbation theory to molecular orbitals \cite{g09}. In these calculations the electric and magnetic dipole moments and the electric quadrupole moment are usually treated as microscopic quantities that apply to single molecules. However, in this paper we discuss the optical activity of a macroscopic medium and, therefore, all the moments included in Eqs. \eqref{moments} should be regarded as macroscopic moment densities, i. e. their statistical average multiplied by the number density. We are also assuming that local field effects are small enough to be neglected. From this point, we  refer always to moment densities instead of molecular moments and to macroscopic fields instead of local fields.

In a nonmagnetic medium the molecular property tensors $\hat{\alpha}_{\alpha\beta}$, $\hat{G}_{\alpha\beta}$ and $\hat{A}_{\alpha\beta\gamma}$ must define the constitutive dyadics of the bianisotropic Eqs. \eqref{constitutivesimplied2} so that the optical response of an oriented molecule or crystal is calculated from them. Following the work of Graham and Raab \cite{raab1} we find the following relations:
\begin{subequations}\label{unodos}
\begin{equation}
(\rtta{\varepsilon})_{\alpha\beta}=\varepsilon_0\delta_{\alpha\beta}+\hat{\alpha}_{\alpha\beta},
\end{equation}
\begin{equation}
(\rtta{\mu})_{\alpha\beta}=\mu_0^{-1}\delta_{\alpha\beta},
\end{equation}
\begin{equation}\label{factor}
(\rtta{\kappa})_{\alpha\beta}=\frac{1}{2}[(-i\hat{G}_{\alpha\beta}-i\hat{G}_{\beta\alpha})+\frac{\omega}{3}(\varepsilon_{\beta\gamma\delta}\hat{A}_{\gamma\delta\alpha}+\varepsilon_{\alpha\gamma\delta}\hat{A}_{\gamma\delta\beta})],
\end{equation}
\end{subequations}
where $\varepsilon_{\beta\gamma\delta}$ is the Levi-Cevita operator and $\delta_{\alpha\beta}$ is the Kronecker delta. We are using the sum over indices Einstein convention and the notation is the same as in Refs. \cite{book2,dunn}. Eq. \ref{factor} can be found, apart from a numerical factor, in Ref. \cite{dunn}, but it is different from its equivalent appearing in Ref. \cite{raab1} because here we have disregarded nonreciprocal factors. As natural optical activity is a reciprocal optical phenomena  only reciprocal contributions (time-even) need to be considered and the nonreciprocal factors (time-odd) can be safely neglected. The condition $\rtta{\rho}=-\rtta{\rho}'^T$  is automatically satisfied because $\hat{\alpha}_{\alpha\beta}$ as well as the sums ($\hat{G}_{\alpha\beta}+\hat{G}_{\beta\alpha}$) and ($\varepsilon_{\beta\gamma\delta}\hat{A}_{\gamma\delta\alpha}+\varepsilon_{\alpha\gamma\delta}\hat{A}_{\gamma\delta\beta}$) are all symmetric contributions. An additional difference compared to Ref. \cite{raab1} is that we get a 1/3 factor in front of the $A_{\gamma\delta\alpha}$ terms, as opposed to the 1/2 factor obtained by Graham and Raab. This change arises because they used a so-called primitive definition of the electric quadrupole moment compared to the more extended traceless definition used for example by Buckingham and Dunn \cite{dunn}. Both, the traceless and primitive definitions, are applicable because it has been shown that they allow for an origin independent description of the theoretical optical activity \cite{dunn,auts,raab2}, i.e. its final expression is not dependent on the arbitrary choice of coordinate origin.

\section{Optical activity in terms of the magnetolectric tensor of gyrotropy}

Eq. \eqref{factor} is specially important because provides the connection between the molecular polarizability tensors and the magnetoelectric tensor. This equation appears in the seminal publication of Buckingham-Dunn (Eq. [19] in Ref. \cite{dunn}) but, as far as we know, this work does not offer a clear interpretation about the meaning of $\kappa_{\alpha\beta}$. Apparently, it was introduced as an accessory equation for their calculation of optical activity but it was not highlighted as a main result of this publication.

The elements of the  magnetoelectric tensor of gyrotropy $\kappa_{\alpha\beta}$ (from this point we omit the dyadic notation) are calculated expanding Eq. \eqref{factor}:
\begin{subequations}\label{kap}
\begin{equation}\label{one}
\kappa_{xx}=-i\hat{G}_{xx}+\frac{\omega }{3}(\hat{A}_{yzx}-\hat{A}_{zyx}),
\end{equation}
\begin{equation}\label{two}
\kappa_{yy}=-i\hat{G}_{yy}+\frac{\omega }{3}(\hat{A}_{zxy}-\hat{A}_{xzy}),
\end{equation}
\begin{equation}
\kappa_{zz}=-i\hat{G}_{zz}+\frac{\omega }{3}(\hat{A}_{xyz}-\hat{A}_{yxz}),
\end{equation}
\begin{equation}
\kappa_{xy}=\kappa_{yx}=-\frac{i}{2}(\hat{G}_{xy}+\hat{G}_{xy})+\frac{\omega }{6}(\hat{A}_{yzy}+\hat{A}_{zxx}-\hat{A}_{zyy}-\hat{A}_{xzx}),
\end{equation}
\begin{equation}
\kappa_{yz}=\kappa_{zy}=-\frac{i}{2}(\hat{G}_{yz}+\hat{G}_{zy})+\frac{\omega }{6}(\hat{A}_{zxz}+\hat{A}_{xyy}-\hat{A}_{xzz}-\hat{A}_{yxy}),
\end{equation}
\begin{equation}
\kappa_{xz}=\kappa_{zx}=-\frac{i}{2}(\hat{G}_{xz}+\hat{G}_{zx})+\frac{\omega }{6}(\hat{A}_{yzz}+\hat{A}_{xyx}-\hat{A}_{zzy}-\hat{A}_{yxx}),
\end{equation}
\end{subequations}
where $\hat{A}_{\alpha\beta\gamma}$ holds the symmetry property $\hat{A}_{\alpha\beta\gamma}=\hat{A}_{\alpha\gamma\beta}$ \cite{book2}. To relate this result with an experiment it is still necessary to investigate the relation between the optical activity measured in a given molecular direction and the components of $\kappa_{\alpha\beta}$. Optical activity is usually expressed in terms of circular birefringence CB (twice the optical rotation) and circular dichroism (CD). Both quantities can be measured  independently and they can be combined into a complex expression given by
\begin{equation}\label{optact}
\mathrm{C}=\mathrm{CB}-i\mathrm{CD}=(\hat{n}_--\hat{n}_+)\frac{\omega l}{c},
\end{equation}
where $\hat{n}_-$ and $\hat{n}_+$ are, respectively, the complex refractive indices for left an right circularly polarized light. $l$ is the pathlength of the medium. A straight calculation of the values of $\hat{n}_-$ and $\hat{n}_+$  from the tensors $\varepsilon_{\alpha\beta}$, $\mu_{\alpha\beta}$ and $\kappa_{\alpha\beta}$ is in general a complicate task. The problem of light propagation (assuming planewaves) through a reciprocal bianosotropic medium is usually treated by eigenanalysis of a wave equation, which provides the refractive indices of the eigenpolarizations that propagate through the medium. For chiral isotropic medium or along the optic axis of a chiral uniaxial medium in which the eigenmodes are circularly polarized waves  it can be easily shown that C depends only on the magnetoelectric parameter or tensor \cite{malaka}.  However,  the two eigenmodes, with complex refractive indices $\hat{n}_{\sigma}$ and $\hat{n}_{\delta}$, in general do not correspond with circularly polarized waves and, therefore, the values of $\hat{n}_-$ and $\hat{n}_+$ are not directly obtained.

With the help of the Jones formalism for polarization optics, we showed in Ref. \cite{recno} that optical activity  ($\hat{n}_--\hat{n}_+$) could be calculated from the complex retardation between the eigenmodes $(\hat{n}_{\sigma}-\hat{n}_{\delta})$  with the following equation:
\begin{equation}\label{mybestequation}
\hat{n}_--\hat{n}_+=(\hat{n}_{\sigma}-\hat{n}_{\delta})i\frac{1+k_{\sigma}k_{\delta}}{k_{\sigma}-k_{\delta}}
\end{equation}
in which $k_{\sigma}$ and $k_{\delta}$ are given by the polarization of each eigenmode, namely $k_{\delta}=E_{\delta\perp}/E_{\delta\parallel}$ and $k_{\sigma}=E_{\sigma\perp}/E_{\sigma\parallel}$, where $E_i$ are electric field amplitudes. Note that in simple case of a chiral isotropic medium the eigenmodes are circularly polarized waves $k_{\sigma}=i$ and $k_{\delta}=-i$ and, obviously, $\hat{n}_--\hat{n}_+=\hat{n}_{\sigma}-\hat{n}_{\delta}$.

Using Eq. \eqref{mybestequation} it is possible to calculate the dependence of optical activity with the material's constitutive tensors of Eq. \eqref{constitutivesimplied2} for any direction of light propagation. In appendix A we show the results of this calculation for an uniaxial medium with crystallographic point group of symmetry 32, 422 or 622 and in which the optic axis is perpendicular to the direction of propagation of light. In practice, for media with lower symmetry it is very difficult to keep the eigenanalysis at an analytical level and it is more practical to do the calculation numerically. For constitutive tensors belonging to any crystal symmetry it is found, either analytically (as in Appendix A) or numerically, that C depends \emph{only} on the components of the tensor magnetoelectric tensor $\kappa_{\alpha\beta}$, so that this tensor fully describes the optical activity of the sample and that $\varepsilon_{\alpha\beta}$ is completely irrelevant to optical activity.

For any crystal class, the dependence of C with $\kappa_{\alpha\beta}$ follows a general rule: C can be calculated from $\kappa_{\alpha\beta}$ by adding the two components of the tensor that are perpendicular to the direction of propagation of the beam. This simple result was already suggested for some particular cases in \cite{recno}, but now we can generalized it with the following equations that apply to ensembles of molecules:
\begin{subequations}\label{orto}
\begin{equation}
C_{xx}=\frac{\omega}{c}\mu_0lN<\kappa_{yy}+\kappa_{zz}>,
\end{equation}
\begin{equation}
C_{yy}=\frac{\omega}{c}\mu_0lN<\kappa_{xx}+\kappa_{zz}>,
\end{equation}
\begin{equation}\label{cb1}
C_{zz}=\frac{\omega}{c}\mu_0lN<\kappa_{xx}+\kappa_{yy}>,
\end{equation}
\begin{equation}
C_{xy}=\frac{\omega}{c}\mu_0lN<\kappa_{zz}+\kappa_{yx}>,
\end{equation}
\begin{equation}
C_{xz}=\frac{\omega}{c}\mu_0lN<\kappa_{yy}+\kappa_{zx}>,
\end{equation}
\begin{equation}
C_{yz}=\frac{\omega}{c}\mu_0lN<\kappa_{xx}+\kappa_{zy}>,
\end{equation}
\end{subequations}
where the angular brackets $<>$ denote a statistical average. $N$ is the number of molecules per unit volume and $l$ is the pathlength. The subscript in $C_{ij}$ indicates the direction of measurement of optical activity, e.g. $C_{xy}$ corresponds to optical activity in the direction that bisects the $x$ and $y$ axes. As $\kappa_{ij}=\kappa_{ji}$, it immediately follows that $C_{ij}=C_{ji}$ because natural optical activity is a reciprocal phenomenon.

A qualitative understanding of the dependence of C with $\kappa_{\alpha\beta}$ is gained by realizing that a circularly polarized wave can be decomposed into the sum of two orthogonal linearly polarized components that are in the plane of polarization and have a phase difference of $90^{\circ}$. The sign of this phase difference determines the handedness of the wave. Only the tensor components that correspond to two these two orthogonal directions are able to contribute in a different way to left- and right- circularly polarized waves. The other non-vanishing tensor elements (either in the magnetoelectric or dielectric tensors) affect the absorption and refraction of light but they have exactly the same effect for left- and right- circularly polarized wave and, therefore, they do not contribute to C.

For example, to calculate the optical activity corresponding to light propagating in the z or -z direction of an oriented medium  Eq.\eqref{cb1} is used in combination with Eqs. \eqref{one} and \eqref{two}:
\begin{equation}
C_{zz}=\frac{\omega}{c}\mu_0lN<-i\hat{G}_{xx}-i\hat{G}_{yy}+\frac{\omega}{3}(\hat{A}_{yzx}-\hat{A}_{xyz})>,
\end{equation}
where it has been used that $A_{zyx}=A_{zxy}$. According to Ref. \cite{dunn} at transparent frequencies $\hat{G}_{\alpha\beta}=-iG'_{\alpha\beta}$ and $\hat{A}_{\alpha\beta\gamma}=A_{\alpha\beta\gamma}$, where $G'_{\alpha\beta}$ and $A_{\alpha\beta\gamma}$ are real tensors, so it is possible to rewrite:
\begin{equation}
CB_{zz}=-\frac{\omega}{c}\mu_0lN<G'_{xx}+G'_{yy}+\frac{\omega}{3}(\hat{A}_{xyz}-\hat{A}_{yzx})>,
\end{equation}
which is the same result found in \cite{dunn}.

\begin{figure}[htb]
\includegraphics[width=17cm]{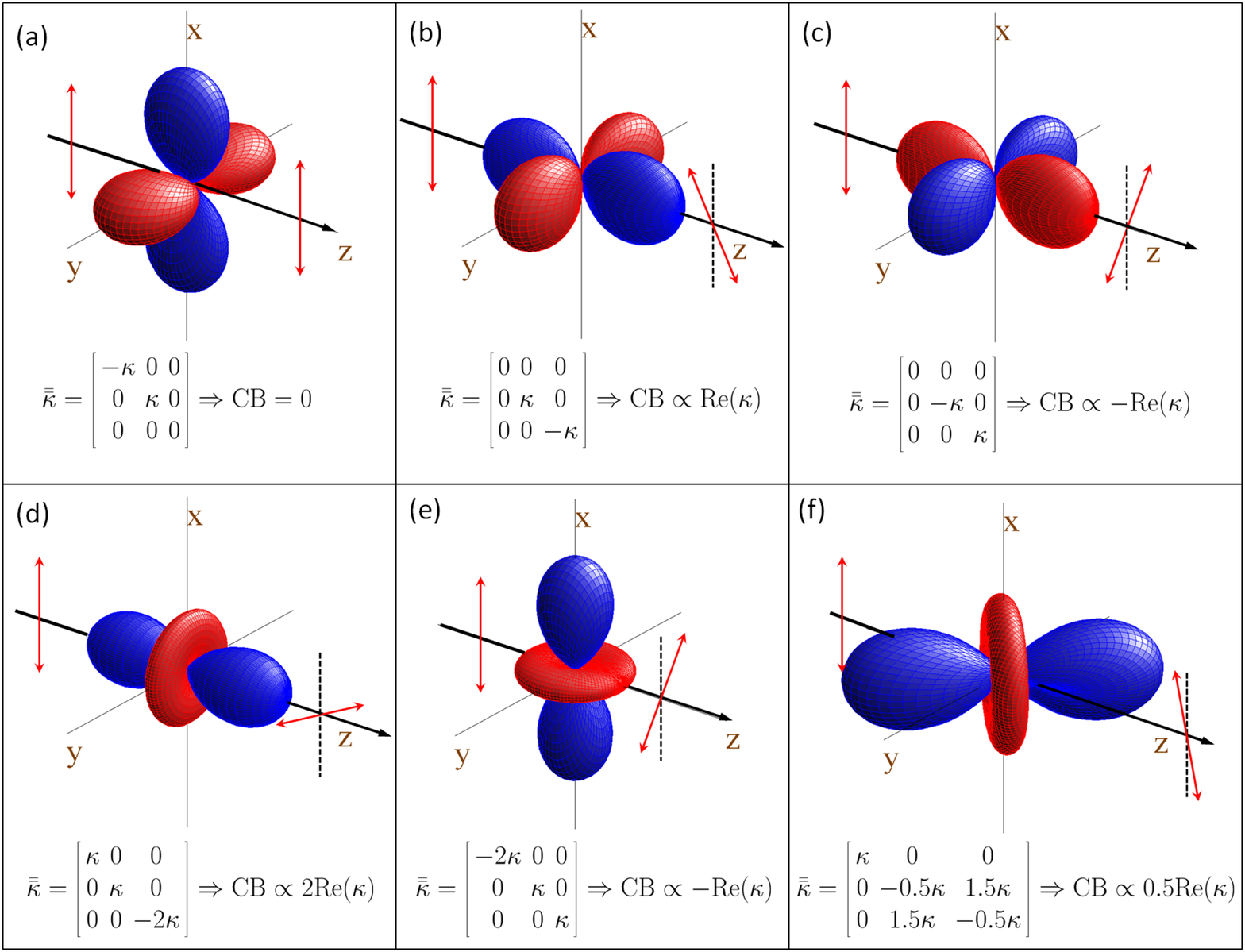}
\begin{quote}
\caption{Visual representation \cite{kam} of the magnetoelectric tensor and the induced circular birefringence on a linearly polarized light beam propagating along the z axis. Red color represents positive components of the magnetoelectric tensor and blue indicates the negatives. Panels a, b and c show different orientations of a molecule with crystallographic point group of symmetry $\bar{4}2m$ or mm2. Panels d, e and f correspond to a trigonal molecule with point group of symmetry 3, 32 or 622 in which the two independent components of its magnetoelectric tensor have a relation -2:1.  In all cases the resulting CB is given by the real part of the sum of the two components of the magnetoeletric tensor that are perpendicular to the direction of propagation: $\kappa_{xx}+\kappa_{yy}$.} \label{figtensor}
\end{quote}
\end{figure}

As a practical demonstration of the implications of Eqs. \eqref{orto}, we have plotted in Fig. \ref{figtensor} a visual representation of the value of CB as a function of the orientation of a molecule. In all cases we have considered an incoming linearly polarized light beam propagating along the z axis. Panels a, b and c correspond to a molecule with crystallographic point group of symmetry $\bar{4}2m$ or mm2 at three different orientations. The water (H$_2$O) molecule could be a good an example of this symmetry, which is characterized by a magnetoelectric tensor having only one independent component.  Molecules belonging to this group do not have optical activity in solution because when all directions of the space are considered the magnetoelectric tensor averages to zero (which happens because the trace of the tensor matrix is always zero) \cite{kaminsky0}. However, for the orientations shown in b and c there exits CB and it, respectively, takes opposite signs. Panels d, e and f show a molecule with crystallographic point group of symmetry 3, 32 or 622 that has a magnetoelectric tensor with two different tensor components. In this example we have considered that these two components keep a relation -2:1. Note that the trace of the magnetoelectric tensor for molecules holding this ratio would be zero and, consequently, they would neither be optically active in solution. For each case the CB along the z axis is given by the real part of the sum of the tensor elements $\kappa_{xx}+\kappa_{yy}$, which is theoretically calculated using Eq. \eqref{cb1}.

\section{Comparison with the classical tensor description for optical activity}

The correspondence between calculated of molecular tensor and optical activity values is treated in other works considering another tensor, usually called gyration tensor or optical activity tensor, and typically represented by $g_{\alpha\beta}$ \cite{newnham,book2,raabcubic,veronica}. This tensor is based in the so-called equation of the normals:
\begin{equation}\label{eqNO}
(n^2-n^2_{01})(n^2-n^2_{02})=G^2,
\end{equation}
where $n$ gives the possible values of the refractive  of the refractive index, for a given direction of the wave normal, $n_{01}$ and $n_{02}$ are the refractive indices of the eigenwaves propagating in the crystal in the absence of optical activity, and $G$ is the scalar gyration parameter
\begin{equation}
G=g_{\alpha\beta}N_{\alpha}N_{\beta}.
\end{equation}
$N_{\alpha}$ and $N_{\beta}$ are direction cosines of the wave normal. Eq. \eqref{eqNO} is typically approximated to $(n^2-\bar{n}^2)=G^2$ by assuming that the birefringence of the system is not too large ($n_{01}\cong n_{02}\cong \bar{n} =\sqrt{n_{01}n_{02}}$). Then if $G$ is assumed to be very small, the two solutions of this equation can be written as $n=\bar{n}\pm G/2n$ that correspond to the refractive indices for left and right circularly polarized waves. Then the optical activity [Eq. \eqref{optact}] at a given direction can be written as
\begin{equation}
C_{ij}=\frac{\omega l}{\bar{n}c}G= \frac{\omega l}{\bar{n}c}g_{\alpha\beta}N_{\alpha}N_{\beta},
\end{equation}
this equation is similar to Eqs. \eqref{orto}, but here the magnetoelectric tensor does not appear at all and and it has been replaced by a different tensor.

The description of optical activity based on the tensor $g_{\alpha\beta}$ is problematic not only because it works uniquely within the approximations that we have detailed above, but also because it arises from an incorrect formulation of the constitutive equations for optical activity. To the best of our knowledge Eq. \eqref{eqNO} was first proposed by Szivessy in 1928 \cite{szi} and it requires constitutive equations in which the permittivity tensor is perturbed by the gyration tensor. This formulation corresponds to the classic treatment of optical activity by M. Born \cite{born}, which neglects any magnetoelectric tensor and optical activity is then justified by modification of the permittivity tensor:
\begin{equation}
\varepsilon_{ik}=\varepsilon_{ik}^{(0)}-i\epsilon_{ikl}g_{lm}N_{m},
\end{equation}
where $\epsilon_{ikl}$ is the Levi-Civita symbol, $g_{lm}$ are the components of the optical activity tensor and $N_{m}$ are the direction cosines of the wave normal. This form of the constitutive equations is similar to the one used for describing magneto-optical phenomena (Faraday effect and magneto-optic Kerr effect) but, despite being quite widespread in classical crystal optics texts \cite{nye,newnham}, it is not suitable for natural optical activity as it violates fundamental principles that natural optical activity should preserve \cite{laka2,silvermangyro}.

The magnetoelectric tensor $\kappa_{\alpha\beta}$ corresponding every crystal class takes the same forms of symmetry as the gyration tensor $g_{\alpha\beta}$  (see for example the tables in Refs. \cite{nye, newnham}) because the same constraints based on Neumann's principle are applicable.  However, the values of the non-vanishing elements of $\kappa_{\alpha\beta}$ bear no relation with those of $g_{\alpha\beta}$. In general, we strongly recommend to use $\kappa_{\alpha\beta}$ instead of $g_{\alpha\beta}$ to express the experimental or calculated optical activity of oriented systems.

\section{Conclusions}

The optical activity of anisotropic crystals or oriented molecules is adequately described by a reciprocal bianisotropic formulation of the constitutive equations. CB and CD for any molecular or crystallographic direction are defined by the symmetric magnetoelectric tensor of chirality $\kappa_{\alpha\beta}$ which, in turn, is expressed in terms of two molecular property tensors. These contain the effect of the mean electric and magnetic dipoles induced, respectively, by the time-derivative of the magnetic and electric fields of the light wave, as well as the electric dipole induced by the electric field gradient of the wave and the electric quadrupole induced by the electric field.

We have shown that the optical activity found by a light plane wave propagating along a certain direction of a molecule or a crystal is given by the sum of two components of $\kappa_{\alpha\beta}$ that are perpendicular to this direction of propagation. These two tensor components describe the change of polarization of light that is due to the special interaction between the electric and magnetic fields in systems with optical activity. With  bianisotropic constitutive equations this change in polarization is only described by the magnetoelectric tensor and it is separable from the change in polarization due to the linear birefringence or linear dichroism typical of oriented systems. We expect that the use of the bianisotropic formalism will simplify the comparison between experiments and calculations and it will bring the field of optical activity in molecules to a closer connection with that of artificial metamaterials.

\section*{Acknowledgements}

The author acknowledges financial support from a  Marie Curie IIF Fellowship (PIIF-GA-2012-330513 Nanochirality). He is also grateful to V. Murphy and B. Kahr for helpful discussions.


%
%

%


\newpage
\appendix

\section{Optical activity in a birefringent direction of the symmetry classes 32, 422 and 622}

This appendix includes an analytical calculation of the dependence of optical activity with the magnetoelectric tensor for a medium with crystallographic symmetry 32, 422 or 622. For this example we assume that light propagates in a direction perpendicular to the optic axis. For a medium with this symmetry, the dielectric tensor is $\rtta{\varepsilon}=\mathrm{diag}(\varepsilon_{11}, \varepsilon_{11}, \varepsilon_{33})$ and the magnetoelectric tensor is $\rtta{\rho}=\mathrm{diag}(i\kappa_{11}, i\kappa_{11}, i\kappa_{33})$.

After performing the eigenalysis of the wave equations using the produce described in Ref.\cite{recno} we find that the refractive indices of the eigenmodes are given by.
\begin{equation}
\hat{n}_{\sigma,\delta}=2^{1/2}[\varepsilon_{11}+\varepsilon_{33}+2\kappa_{11}\kappa_{33}\mp\xi ]^{1/2}/2,
\end{equation}
in which $\xi=\sqrt{(\varepsilon_{33}-\varepsilon_{11})^2+4(\kappa_{11}+\kappa_{33})(\kappa_{11}\varepsilon_{33}+\kappa_{33}\varepsilon_{11})}$.

$k_{\sigma}$ and $k_{\delta}$ are obtained from the ratio of the electric field amplitudes appearing in the components of the eigenvectors associated to these eigenvalues:
\begin{subequations}
\begin{equation}
k_{\sigma}=-\frac{[\kappa_{11}\varepsilon_{11}+2\kappa_{33}\varepsilon_{11}+\kappa_{11}\varepsilon_{33}-\kappa_{11}\xi]i}{(\varepsilon_{33}-\varepsilon_{11}+\xi)\hat{n}_{\sigma}},
\end{equation}
\begin{equation}
k_{\delta}=\frac{[\kappa_{11}\varepsilon_{11}+2\kappa_{33}\varepsilon_{11}+\kappa_{11}\varepsilon_{33}+\kappa_{11}\xi]i}{(\varepsilon_{11}-\varepsilon_{33}+\xi)\hat{n}_{\delta}}.
\end{equation}
\end{subequations}

When all these equations are combined according to Eq. \eqref{mybestequation} there are many simplifications and the final result is extremely simple. The optical activity for this direction of light propagation is given by:
\begin{equation}
\hat{n}_--\hat{n}_+=(\hat{n}_{\sigma}-\hat{n}_{\delta})i\frac{1+k_{\sigma}k_{\delta}}{k_{\sigma}-k_{\delta}}=\kappa_{11}+\kappa_{33}.
\end{equation}
Case e of Fig. 1 shows a pictorial representation of this particular relationship between optical activity and the magnetoelectric tensor.

\newpage

\end{document}